\documentclass[conference]{IEEEtran}
\IEEEoverridecommandlockouts
\usepackage{cite}
\usepackage{amsmath,amssymb,amsfonts}
\usepackage{algorithmic}
\usepackage{graphicx}
\usepackage{textcomp}
\usepackage{xcolor}
\def\BibTeX{{\rm B\kern-.05em{\sc i\kern-.025em b}\kern-.08em
    T\kern-.1667em\lower.7ex\hbox{E}\kern-.125emX}}
\begin{document}

\title{PTracer: A Linux Kernel Patch Trace Bot \\
\thanks{\textsuperscript{*}These authors contributed equally to this work.}
\thanks{\textsuperscript{1}CGEL is ZTE embedded operating system based on Linux kernel. It won the China Industry Award in 2016 and the Gold Award of China International Software Expo in 2017.}
}

\author{\IEEEauthorblockN{Yang Wen\textsuperscript{*} \qquad Jicheng Cao\textsuperscript{*} \qquad Shengyu Cheng\textsuperscript{*}}
\IEEEauthorblockA{\textit{Department of Operating System Product} \\
\textit{ZTE Corporation}\\
Chengdu, China \\
\{wen.yang99, cao.jicheng1, cheng.shengyu\}@zte.com.cn}
}

\maketitle

\begin{abstract}
We present PTracer, a Linux kernel patch trace bot based on an improved PatchNet. PTracer continuously monitors new patches in the git repository of the mainline Linux kernel, filters out unconcerned ones, classifies the rest as bug-fixing or non bug-fixing patches, and reports bug-fixing patches to the kernel experts of commercial operating systems. We use the patches in February 2019 of the mainline Linux kernel to perform the test. As a result, PTracer recommended 151 patches to CGEL kernel experts out of 5,142, and 102 of which were accepted. PTracer has been successfully applied to a commercial operating system and has the advantages of improving software quality and saving labor cost.
\end{abstract}


\section{Introduction}\label{INTRO}
It is very important for the maintainers of enterprise commercial operating systems based on Linux kernel to monitor the open source patches and pick bug-fixing ones. They mainly face the following challenges: (1) It is very difficult to manually identify patches that should be merged into the stable versions because of massive patches. (2) Maintainers of the stable versions may miss some important patches due to omission when selecting patches. (3) Commercial operating systems are usually tailored for specific products, the patches for unrelated modules should be ignored.

As is described in \cite{b1}, neither keyword-based nor Bugzilla-based approaches are sufficient to meet these challenges. PatchNet \cite{b2} demonstrates a promising prospect of using machine learning to classify Linux kernel patches, however, as a research demo, it can not be directly applied to engineering applications because of the following limitations. (1) The training and the evaluation data set are fixed, the trained model can not be improved continuously for lack of feedback. (2) The trained PatchNet model cannot be directly deployed and used for prediction, because the preprocessing process does not save the mapping relationship between the text format patch file and the digital format intermediate file. When predicting for new patches, the new patches must be reprocessed with all original patches in the training and the evaluation data set to generate a new intermediate file, and an extra retraining is needed. This is a time comsuming process. (3) PatchNet can partly solve the first two challenges mentioned above, and it can not solve the third challenge at present.

Based on an improved PatchNet, we constructed PTracer, which can continuously monitor new patches submitted to the mainline Linux kernel, classify bug-fixing ones, and report recommended patches to kernel experts. Our contributions are:

\begin{itemize}
\item A closed-loop trace bot of Linux kernel patches is developed.
\item A number of engineering improvements for the preprocessing approaches and the neural network of PatchNet have been made, which brings out a practical use in commercial operating systems.
\item We have evaluated PTracer on CGEL\textsuperscript{1} and show that it can improve software quality and save labor cost.
\end{itemize}

\section{Overview of PTracer}

\begin{figure}[b]
\centerline{\includegraphics[width=2.5in]{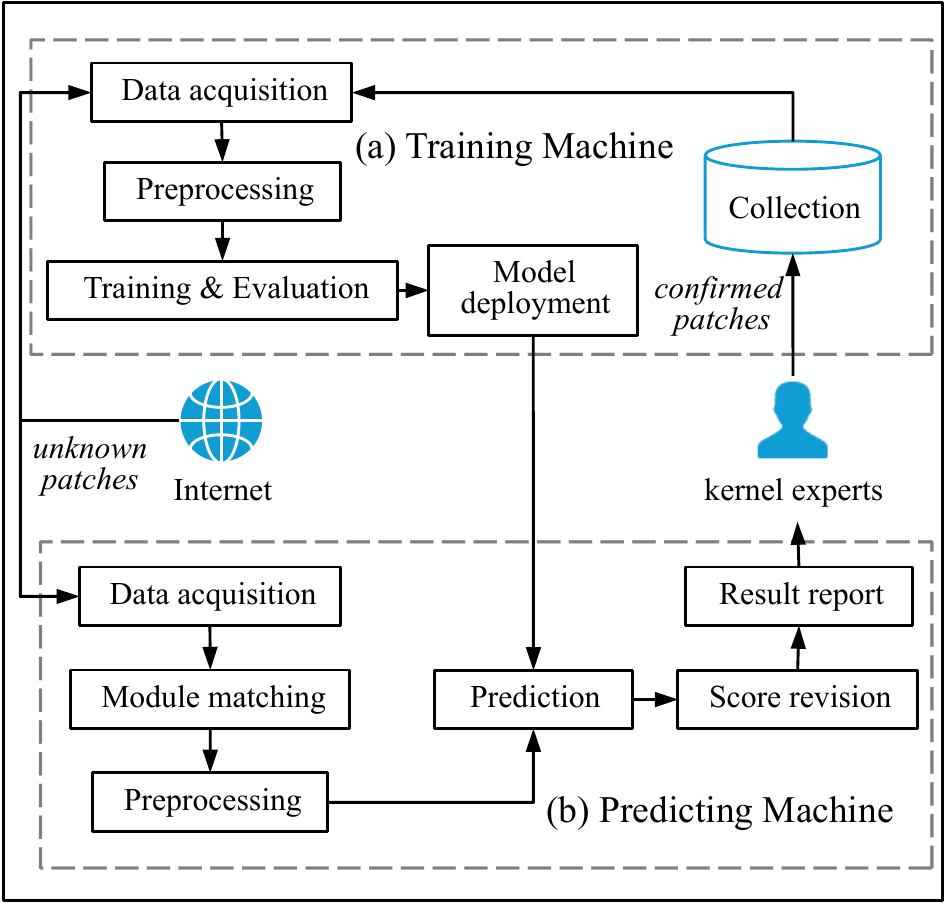}}
\caption{The overview workflow of PTracer.}
\label{fig1}
\end{figure}

The overall framework of PTracer is shown in Fig.~\ref{fig1}. It contains three stages: training, predicting and feedback. 

The training stage consists of four modules, which work on the Training Machine. The data acquisition module collects patches from the git repository of the mainline Linux kernel and labels each patch as bug-fixing or non bug-fixing. The patches confirmed by kernel experts of CGEL will be also considered. The preprocessing module builds a vocabulary for all patches. The patches will be marshalled into a digital format intermediate file as the input to the neural network. The training \& evaluation module trains and evaluates the neural network repeatedly until we get a satisfactory model. The model deployment module will package and deploy the trained model with its dependencies.

The predicting stage works on the Predicting Machine. The data acquisition module continuously monitors new patches in the git repository of the mainline Linux kernel, the monitoring period can be specified as one day or longer. We maintain a concerned module list (e.g. arch/x86) of CGEL to filter out the patches that does not belong to the concerned modules by the module matching. The preprocessing module in the predicting stage only loads the saved vocabulary and preprocess patches to be predicted. The prediction module will give each patch a score. The revision module will boost the scores of patches containing some specfic strings, such as “cc: stable@vger.kernel.org”. If the final score is higher than a threshold value, the corresponding patch will be reported to kernel experts of CGEL by the result report module. 

The feedback stage also works on the Training Machine. The kernel experts of CGEL will check every recommended patch, and give a feedback on two points: (1) whether they accept this patch; (2) if not, give the reason. Confirmed patches will be added to the data collection, with which we will retrain PTracer periodically for continuous improvement.

\section{Experiments and Results}
We have two data sets, one is provided by PatchNet, which contains 42,408 bug-fixing patches and 39,995 non bug-fixing patches. The other is provided by PTracer, which includes 29,619 bug-fixing patches and 39,963 non bug-fixing patches. We experimented on a Dell PowerEdge R930 server with Intel Xeon E7-4850 CPUs (128 processors) and 256 GB memory.

\subsection{Experiment on Different Data Sets}
The accuracy of PTracer trained with our data set was 13.7\% higher than that of the data set provided by PatchNet. We found that the difference of data sets may have a significant impact on the performance of the models. 

\subsection{Experiment on Comparison between PatchNet and PTracer}
The accuracy and precision of PTracer without “ccstable” was about 3.8\% higher than that of PatchNet, while the recall was about 10.6\% increased. The main reason may be that PTracer supports multiple changed files in a patch.
In another experiment that we studied the impact of extracting “ccstable” feature in the score revision module. The results demonstrated that the accuracy, the precision, and the recall were all imporved, especially the recall was incresed by 12.9\%.

\subsection{Engineering Application}
\begin{figure}[t]
\centerline{\includegraphics[width=2.5in]{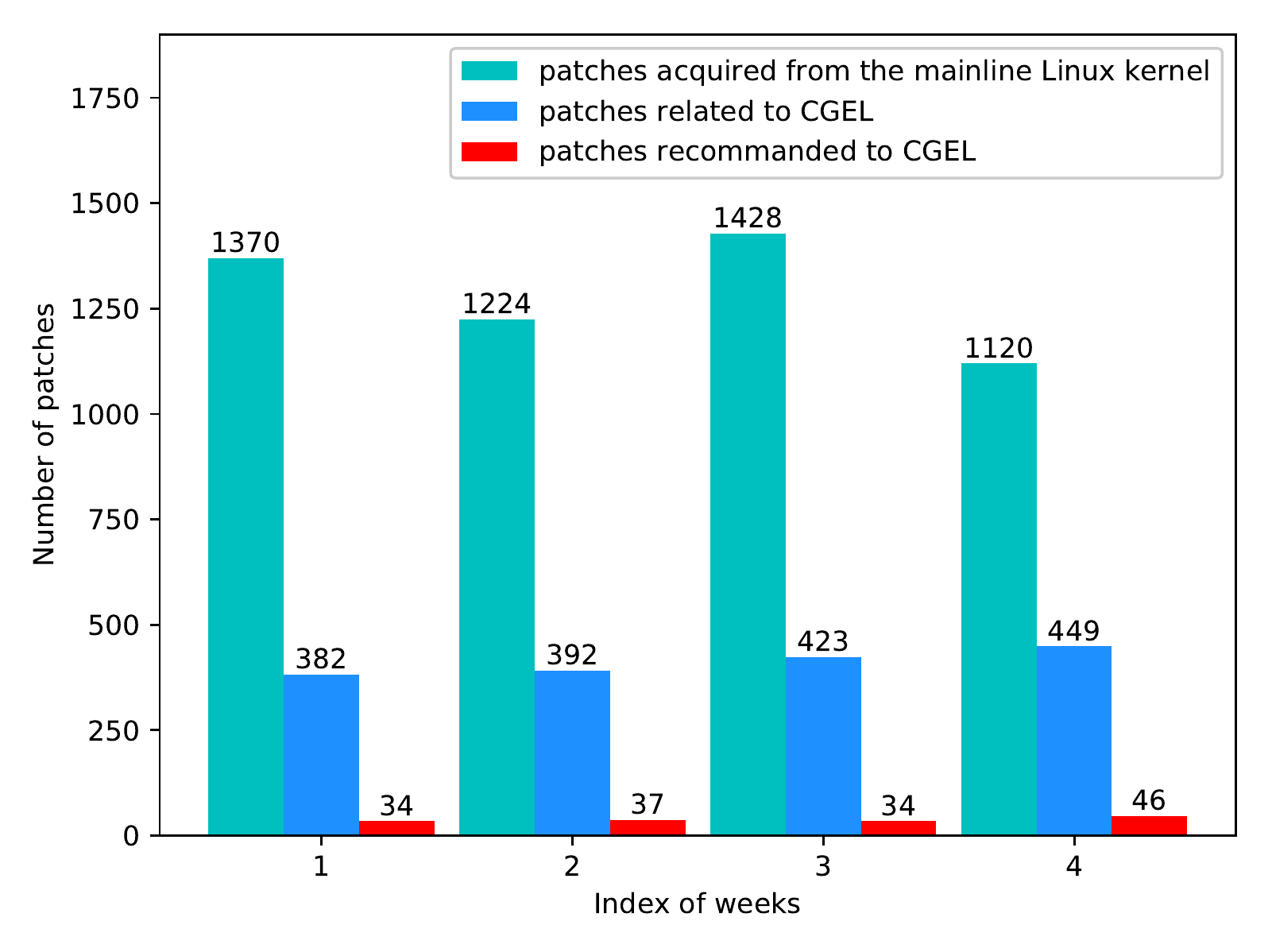}}
\caption{Test results of PTracer with patches in Feb. 2019 of Linux kernel.}
\label{fig2}
\end{figure}

Finally, we applied PTracer with “ccstable” to CGEL and tested it with the patches of mainline Linux kernel between February 1, 2019 and February 28, 2019. The results are shown in Fig.~\ref{fig2}, totally 5,142 patches were analyzed, 1,646 of them were related to CGEL, and 151 patches were recommended to kernel experts. Eventually, 102 out of 151 patches were accepted by CGEL. Among the 49 unaccepted patches, 33 were non bug-fixing, 7 were unrelated bug-fixing patches mistakenly recommended to CGEL due to the coarse granularity of the concerned module list, 6 were not relevant to the baseline version of CGEL, 2 were dependent on some other unincorporated patches, and 1 was for other reasons.

Because of the limitations of PatchNet mentioned in Section \ref{INTRO}, we can not compare the final recommendations by PTracer with those by PatchNet.

\section{Related Works}
In the early stages of patch identification, researchers pick bug-fixing patches through a number of keyword-based approaches \cite{b3}, \cite{b4}. They may miss many important patches for lack of specific keywords. As further improvements, Lawall et al. \cite{b5}, \cite{b6} and Tian \cite{b7} used machine learning to distinguish patches that fix bugs from others, and Hoang et al. \cite{b2} proposed an automated tool named PatchNet. We proposed the PTracer based on an improved PatchNet.

\section{Conclusion}
We built a Linux kernel patch trace bot, and successfully applied it to CGEL, the results were quite encouraging. PTracer has the advantages of improving software quality as well as saving labor cost. In the future, we plan to further improve PTracer's performance, and enhance it to support some other software written in C or other programming languages.

\end{document}